\documentclass[12pt,a4paper,oneside]{article}
\usepackage{epsfig}
\usepackage{pstcol}
\def\PL #1 #2 #3 {{\it Phys.~Lett.~} {\bf#1} (#3) #2}
\def\NP #1 #2 #3 {{\it Nucl.~Phys.~} {\bf#1} (#3) #2}
\def\ZP #1 #2 #3 {{\it Z.~Phys.~} {\bf#1} (#3) #2}
\def\PRL #1 #2 #3 {{\it Phys.~Rev.~Lett.~} {\bf #1} (#3) #2}
\def\PR #1 #2 #3 {{\it Phys.~Rev.~} {\bf#1} (#3) #2}
\def\MPL #1 #2 #3 {{\it Mod.~Phys.~Lett.~} {\bf#1} (#3) #2}
\def\RMP #1 #2 #3 {{\it Rev.~Mod.~Phys.~} {\bf#1} (#3) #2}
\def\EPJ #1 #2 #3 {{\it Eur.~Phys.~J.~} {\bf#1} (#3) #2}

\def\ifm{\ifmmode}

\definecolor{Pink}{rgb}{1.,0.75,0.8}
\newcommand{\beq}{\begin{equation}}
\newcommand{\eeq}{\end{equation}}
\newcommand{\beqn}{\begin{eqnarray}}
\newcommand{\eeqn}{\end{eqnarray}}
\newcommand{\beqs}{\begin{eqnarray*}}
\newcommand{\eeqs}{\end{eqnarray*}}

\newcommand{\pdf}{PDF\ }
\newcommand{\pdfs}{PDF's\ }

\newcommand{\pdfsb}{PDF's}
\newcommand{\W}{$W$-boson\ }                   
\newcommand{\Z}{$Z$-boson\ }                   
\newcommand{\F}{\mbox{$\cal F$}}

\begin{document}
\thispagestyle{empty}
\begin{center}
{\Large Hard Scattering Based Luminosity Measurement at Hadron Colliders}
\vskip 1cm
Walter T.~Giele\\
{Fermi National Accelerator Laboratory, Batavia, IL 60510} \\
\vskip .4cm
{\bf and}\\
\vskip .4cm
St\'ephane A.~Keller\\
{Theory Division, CERN, CH 1211 Geneva 23, Switzerland} 
\footnote{Supported by the European Commission under
    contract number ERB4001GT975210, TMR - Marie Curie Fellowship}\\ 
\vskip .6cm
\end{center}
\vskip .3cm
\begin{abstract}
A strategy to determine the luminosity at  Hadron Colliders
is discussed using the simultaneous \W and \Z event counts.
The emphasis of the study
will be on the uncertainty induced by the parton density functions.
Understanding this source of uncertainties is crucial for a reliable
luminosity determination using the \W and \Z events.
As an example we will use the D0 run 1 results to extract the
luminosity using the vector boson events and compare the result with the traditional
method. Subsequently we will look at the implications for the top cross section
uncertainties using the extracted luminosity.  

\end{abstract}
\newpage
\section{Introduction}

A luminosity measurement based on a well understood hard
scattering process is desirable. Such a method gives good
control over the theoretical uncertainties and a systematic
approach to further reduce the uncertainties is possible. 
Also, the measured luminosity will be correlated with other hard
scattering processes in the same experiment. This leads
to a smaller uncertainty in the comparison between
experiment and theory as the correlated luminosity uncertainty
partly cancels. Only when comparing results between different
experiments is the full luminosity uncertainty relevant.

The method to determine the luminosity outlined 
in this paper is based on the principle 
of comparing the theoretical
cross section to the measured number of \W events~\cite{Wlumi}. 
However, because of the presence of the \pdf uncertainties the theoretical
prediction is a probability density and
a more sophisticated formalism to extract the luminosity is needed.
Furthermore, by looking at the correlated \W and \Z events simultaneously
we not only measure the luminosity but also provide a consistency check. 
This because the ratio of the \W over the \Z cross sections is independent of
the luminosity.

In section 2 we will review some of the theoretical considerations
needed for the calculations. In particular the use of the optimized
\pdf sets of ref.~\cite{OptimizedPDF} together with the needed
physics parameters used in the predictions of the \W and \Z cross
sections.

Before extracting the luminosity we will first look 
in section 3 at the published D0 \W and \Z
cross sections~\cite{WZRD01a,WZRD01b}. 
Comparing the measured cross sections with the theory predictions, 
which now include the \pdf uncertainties, 
will give us a better understanding of some of the issues involved.

Section 4 will outline the method
and as an example use the D0 run 1a~\cite{WZRD01a} and run 1b~\cite{WZRD01b} 
results to determine the luminosity.
Next, in section 5 we will look at the top quark pair predictions in relation to the
measured luminosity.
First of all we want to predict the measured cross section which
can be compared to other experiments. Secondly, we want to compare 
the measured number of topquark pair events to the theory. In the latter
comparison the luminosity uncertainty is strongly reduced and potentially
challenging the theory further than currently is possible. 

Section 6 summarizes our findings and outlook for
future hadron collider experiments

\section{Theoretical Considerations}
\begin{table}[t]\hspace{-1.75cm}
\begin{tabular}{|c|c||c|c||c|}\hline
$M_W$ (GeV) & $B(W\rightarrow l^\pm\nu)$ & 
$M_Z$ (GeV)& $B(Z\rightarrow l^+l^-)$ & $\alpha^{-1}_{QED}(M_Z)$ \\ \hline\hline
$80.419\pm 0.056$ & ($10.56\pm 0.14$)\% & 
$91.1882\pm 0.0022$ & ($3.3688\pm 0.0026$)\% & 
$128.896\pm 0.090$ \\
\hline\end{tabular}
\caption[]{The value of the physics parameters used in the theory predictions. Their
uncertainties have been neglected with respect to the larger \pdf uncertainties in the
predictions.}
\end{table}

The most important aspect of the method is to be able to quantify
the dominant source of uncertainty in the theory prediction of the
\W and \Z cross sections. The physics parameters used in the prediction
(such as the vectorboson mass and width, the electroweak coupling
constants, etc) are known up to a high precision relative to the experimental
uncertainties. The values used are listed in table~1. However,
the \pdfs carry a large uncertainty incurred by the experimental data
used to determine the \pdfsb. We will use the optimized \pdf sets 
of ref.~\cite{OptimizedPDF}. These \pdf sets have been optimized
with respect to deep inelastic proton scattering data. As we will see
the employed method of numerical integration over the functional space of all
possible \pdfs is well fitted to handle the uncertainty estimates
in the cross section calculations. Important issues such as the correlation
between the \W and \Z cross section predictions induced by the \pdfs and
the non-gaussian aspect of the predictions can be handled without any effort.

The cross section predictions will be performed at next-to-leading order
in the strong coupling constant using the DYRAD Monte Carlo~\cite{DYRAD}. 
While the next-to-next-to-leading order
matrix elements are known~\cite{Neerven}, 
the \pdf evolution is not known up to the matching
order. Moreover, we want to use the extracted luminosity to predict the
number of events for other observables. To be consistent in such a procedure
all theoretical predictions should be at the same order. We can use
the next-to-next-to-leading order matrix element
calculation to get an estimate of the remaining
uncertainties due to the truncation of the perturbative expansion. For the
\W and \Z cross sections this uncertainty is around 2\%
and is well below current experimental uncertainties.
We will also make next-to-leading order cross section predictions for the
topquark pair production using the HVQ Monte Carlo of ref.~\cite{hvq}. This
Monte Carlo is based on the calculations of ref.~\cite{topXcalc}.

A word of caution has to be given to the acceptance corrections
needed for the \W and \Z events given the incomplete leptonic coverage
of the detector. These acceptance corrections have to be calculated 
using the theory model and hence are dependent on the parton density
functions. While this can be easily incorporated in a full analysis
the current published results for the \W, \Z and top quark pair 
production cross sections use a particular parton density function
for the calculation of the acceptance corrections. 
For this paper we have to neglect
this effect which most likely is small compared to other uncertainties 
in the problem. However, it can introduce a bias and only the
experiments themselves could properly take the correlation of the
acceptance correction with the parton density functions into account.

\section{Cross Section Results and Comparisons}
\begin{table}[t]
\begin{center}\begin{tabular}{|l||c||c|}\cline{2-3}
\multicolumn{1}{c|}{}& D0 1a & D0 1b \\ \hline
$\sigma_W$ (nb)&
2.36$\pm$0.02$\pm$0.08$\pm$0.13&
2.31$\pm$0.01$\pm$0.05$\pm$0.10
\\
$\sigma_Z$ (nb)&	
0.218$\pm$0.008$\pm$0.008$\pm$0.011&
0.221$\pm$0.003$\pm$0.004$\pm$0.010
\\
$R$& 
10.82$\pm$0.41$\pm$0.35&
10.43$\pm$0.20$\pm$0.10
\\
${\cal L}^{exp}$ (pb$^{-1}$)&  
12.8$\pm$0.7&
84.5$\pm$3.7
\\
\hline\end{tabular}\end{center}
\caption[]{The D0 run 1a/1b results \cite{WZRD01a,WZRD01b}
for the \W and \Z cross sections specifying the statistical, systematic and
luminosity uncertainty. Also given the ratio $R$ 
with the statistical and systematic uncertainties
and the integrated luminosity ${\cal L}$ measurement.}
\end{table}
\begin{table}[t]
\begin{center}\begin{tabular}{|l||c||c|}\cline{2-3}
\multicolumn{1}{c|}{}& $\sigma(W)$ (nb) & $\sigma(Z)$ (nb)\\ \hline
D0 1a & 2.36+0.15-0.15 & 0.218+0.016-0.016 \\
D0 1b & 2.31+0.11-0.11 & 0.221+0.011-0.011 \\
\hline\hline
MRS99 & 2.49 & 0.218 \\
CTEQ5M & 2.55 & 0.222 \\
\hline\hline
\scriptsize{ZEUS-MRST}              & 2.45+0.06-0.06 & 0.227+0.007-0.007 \\
\scriptsize{NMC-MRST}               & 2.35+0.11-0.09 & 0.231+0.008-0.011 \\
\scriptsize{H1-MRST}                & 2.10+0.18-0.13 & 0.195+0.013-0.013 \\
\scriptsize{H1+LEP-MRST}            & 2.12+0.13-0.16 & 0.194+0.013-0.013 \\
\scriptsize{BCDMS-MRST}             & 2.41+0.12-0.08 & 0.231+0.011-0.007 \\
\scriptsize{BCDMS+LEP-MRST}         & 2.50+0.09-0.11 & 0.237+0.008-0.011 \\
\scriptsize{E665-MRST}              & 2.34+0.09-0.16 & 0.227+0.005-0.019 \\
\scriptsize{E665+LEP-MRST}          & 2.41+0.06-0.18 & 0.227+0.025-0.015 \\
\scriptsize{H1+BCDMS-MRST}          & 2.48+0.09-0.05 & 0.234+0.004-0.008 \\
\scriptsize{H1+BCDMS+LEP-MRST}      & 2.22+0.06-0.12 & 0.208+0.006-0.007 \\
\scriptsize{H1+BCDMS+E665-MRST}     & 2.44+0.07-0.07 & 0.232+0.007-0.006 \\
\scriptsize{H1+BCDMS+E665+LEP-MRST} & 2.35+0.04-0.03 & 0.220+0.006-0.004 \\
\hline
\end{tabular}\end{center}
\caption[]{The 31.73\% confidence level intervals for the \W and \Z cross sections.}
\end{table}

In this section we will look at the quoted D0 \W and \Z cross sections
which uses the nondiffractive inelastic $p\bar{p}$ collision based luminosity 
measurement used by D0. We will
compare the individual \W and \Z results to the theoretical predictions, now including
the \pdf uncertainty. All the experimental results needed in this paper
are collected in table~2. 

Using the DYRAD Monte Carlo with the parameter choice of table~1
we make 100 predictions for the \W and \Z cross sections
for each of the optimized \pdf sets. The 100 \pdfs randomly selected out of a set
of 100,000 \pdfs are sufficient for the analysis in this paper.
This leaves us with the basic probability density function for
the vector boson cross section
\beq\label{theoryD}
P_{pdf}(\sigma_V) = \frac{1}{N}\sum_{i=1}^N \delta(\sigma_V-\sigma_V^t(\F_i))\ ,
\eeq
where the sum runs over the $N=100$ optimized \pdfs in the set.
The theoretical prediction $\sigma_V^t(\F_i)$ depends on the \pdf $\F_i$.
This is a scatter plot representation of the probability density function.
To calculate confidence level intervals based on only the theory prediction
we use a histogram representation of the probability density function
\beq\label{theoryH}
P_{pdf}(\sigma_V) = \frac{1}{\Delta N}\sum_{i=1}^N 
\Theta(\sigma_V-\frac{1}{2}\Delta-\sigma_V^t(\F_i))\times 
\Theta(\sigma_V^t(\F_i)-\sigma_V-\frac{1}{2}\Delta)\ ,
\eeq
with the bin width $\Delta$ to be chosen 0.1 nb for the \W cross section and 0.01 nb for the 
\Z cross section. Using the histogram representation we can define the confidence level probability
\beq\label{CLdef}
CL(\sigma_V)=\int d\,\sigma_V^\prime\ \Theta(L^2(\sigma_V^\prime)-L^2(\sigma_V))
\times P_{pdf}(\sigma_V^\prime)\ , 
\eeq
with the log-likelyhood given by
\beq
L^2(\sigma_V)=-2\times\log(P_{pdf}(\sigma_V)/P_{expmax})\ ,
\eeq
and
\beq
P_{expmax}=\max_{\sigma_V} P_{pdf}(\sigma_V)\ .
\eeq 

\begin{table}[t]\label{XCLexp}
\begin{center}\begin{tabular}{|l||c|c||c|c|}\cline{2-5}
\multicolumn{1}{c|}{}& 
\multicolumn{2}{|c||}{D0 1a} & \multicolumn{2}{c|}{D0 1b} \\ \cline{2-5}
\multicolumn{1}{c|}{}& 
$\sigma(W)$ & $\sigma(Z)$ & $\sigma(W)$ & $\sigma(Z)$ \\ \hline
ZEUS-MRST & 51.4 & 57.6 & 23.4 & 64.1 \\
NMC-MRST & 99.9 & 51.7 & 75.1 & 53.0 \\
{\red H1-MRST} & 18.2 & 22.6 & 23.6 & {\red 8.8} \\
{\red H1+LEP-MRST} & 19.6 & 13.2 & 26.8 & {\red 3.4} \\
BCDMS-MRST & 71.5 & 44.9 & 45.3 & 45.2 \\
BCDMS+LEP-MRST & 48.2 & 34.5 & 23.5 & 30.9 \\
E665-MRST & 85.1 & 84.6 & 94.7 & 95.9 \\
E665+LEP-MRST & 98.6 & 62.7 & 73.3 & 73.4 \\
H1+BCDMS-MRST & 35.3 & 38.5 & 12.5 & 34.7 \\
H1+BCDMS+LEP-MRST & 35.2 & 60.2 & 43.6 & 32.2\\
H1+BCDMS+E665-MRST & 55.7 & 37.4 & 26.1 & 33.9 \\
H1+BCDMS+E665+LEP-MRST & 100 & 81.1 & 67.3 & 95.0 \\
\hline\end{tabular}\end{center}
\caption[]{The confidence levels (in percentages) for the measured values
of the vectorboson cross sections using all optimized PDF sets.}
\end{table}
The results for the 12 optimized \pdfs are shown in table~4 together with the
MRS99~\cite{MRS99} and CTEQ5~\cite{CTEQ5} predictions. 
The predicted one sigma standard deviation
uncertainty varies between 1\% and 8\% depending on the chosen set.

Comparison with the experimental results should be done slightly
different than by looking at overlaps in the confidence level intervals.
The reason is that the experimental response function 
$P_{exp}(\sigma_V^e|\sigma_V^t)$ can be used. This probability density function
gives the probability of measuring $\sigma_V^e$ given a true nature value 
$\sigma_V^t$ and is a condensation of the experimental uncertainty analysis.
In this case the experimental response function is simply a one-dimensional
gaussian with a width equal to the combined statistical and systematic
uncertainties as given in table~2. We no longer have to construct the histogram
of eq.~\ref{theoryH} with an arbitrary parameter $\Delta$. Instead
the \pdf probability density for measuring $\sigma_V^e$ given a particular
\pdf set is  
\beq\label{expH}
P_{pdf}(\sigma_V^e) = \frac{1}{N}\sum_{i=1}^N 
P_{exp}\left(\sigma_V^e|\sigma_V^t(\F_i)\right)\ .
\eeq

Using eq.~\ref{CLdef} we calculate the confidence level of the particular D0
measurements $CL(\sigma_V^{meas})$. That is, the likelihood that a repeat
of the experiment renders a worse agreement with the theory. The results
for both the D0 run 1a and run 1b are given in table~5. As is
clear from the table the agreement with the theory for the run 1a results
is excellent, with the {\tt H1-MRST} and {\tt H1+LEP-MRST} sets 
being the most disagreeing. The comparison
with run 1b is more challenging as the accuracy of the experimental results
was increased dramatically. 
Yet, with the exception of the two {\tt H1} set predictions
all other \pdfs render excellent agreement.

\section{Luminosity Determinations}
\begin{figure}[p]
\label{CrossScatter}
\begin{center}\vspace{17cm}
\includegraphics{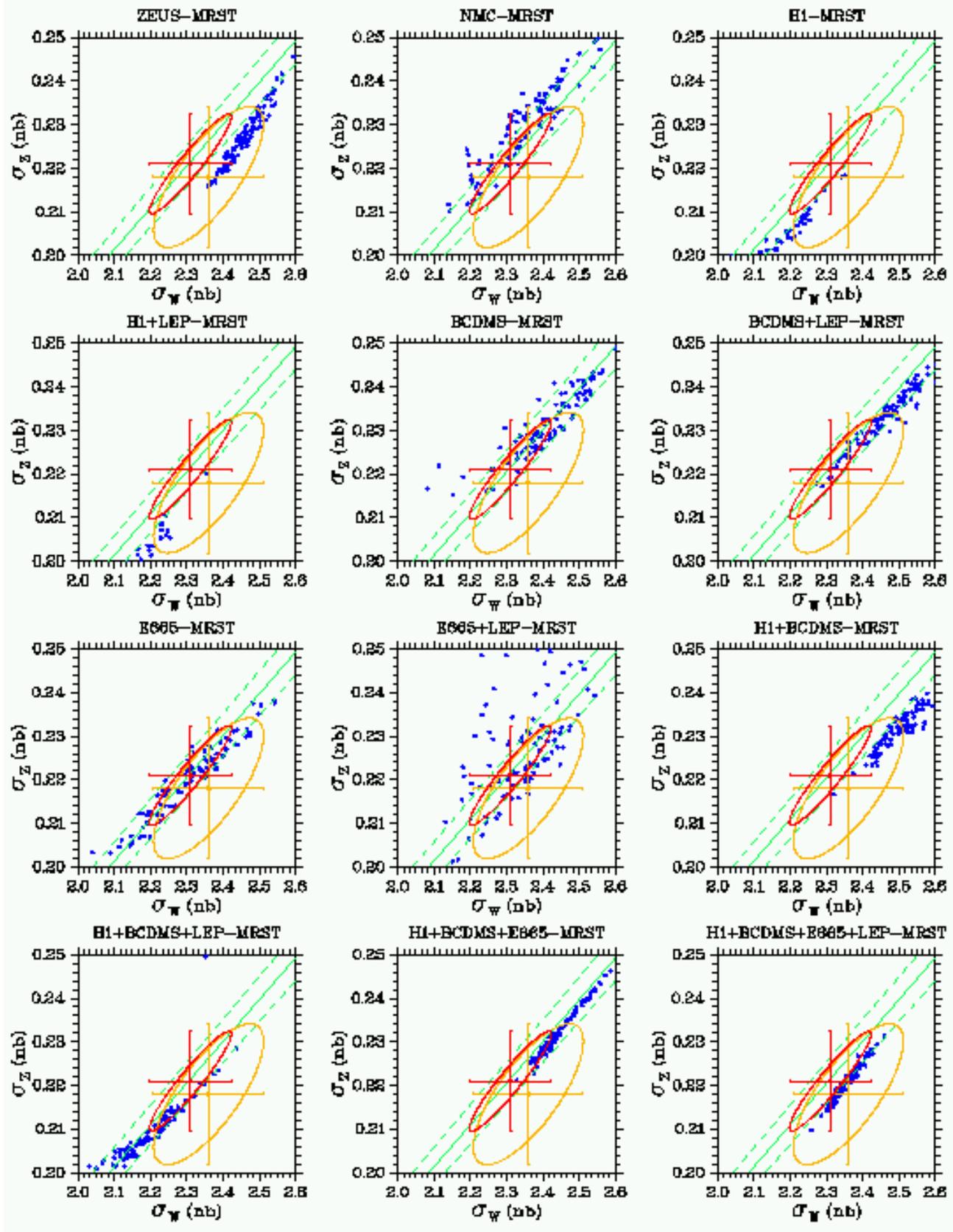}
\end{center}
\caption[]{The optimized \pdf scatter predictions (blue) compared to the D0 run 1a
(magenta) and run 1b (red) measurements. Also indicated is the D0 ratio
measurement (green).}
\end{figure}
\begin{figure}[p]
\label{Luminosity}
\begin{center}\vspace{17cm}
\includegraphics{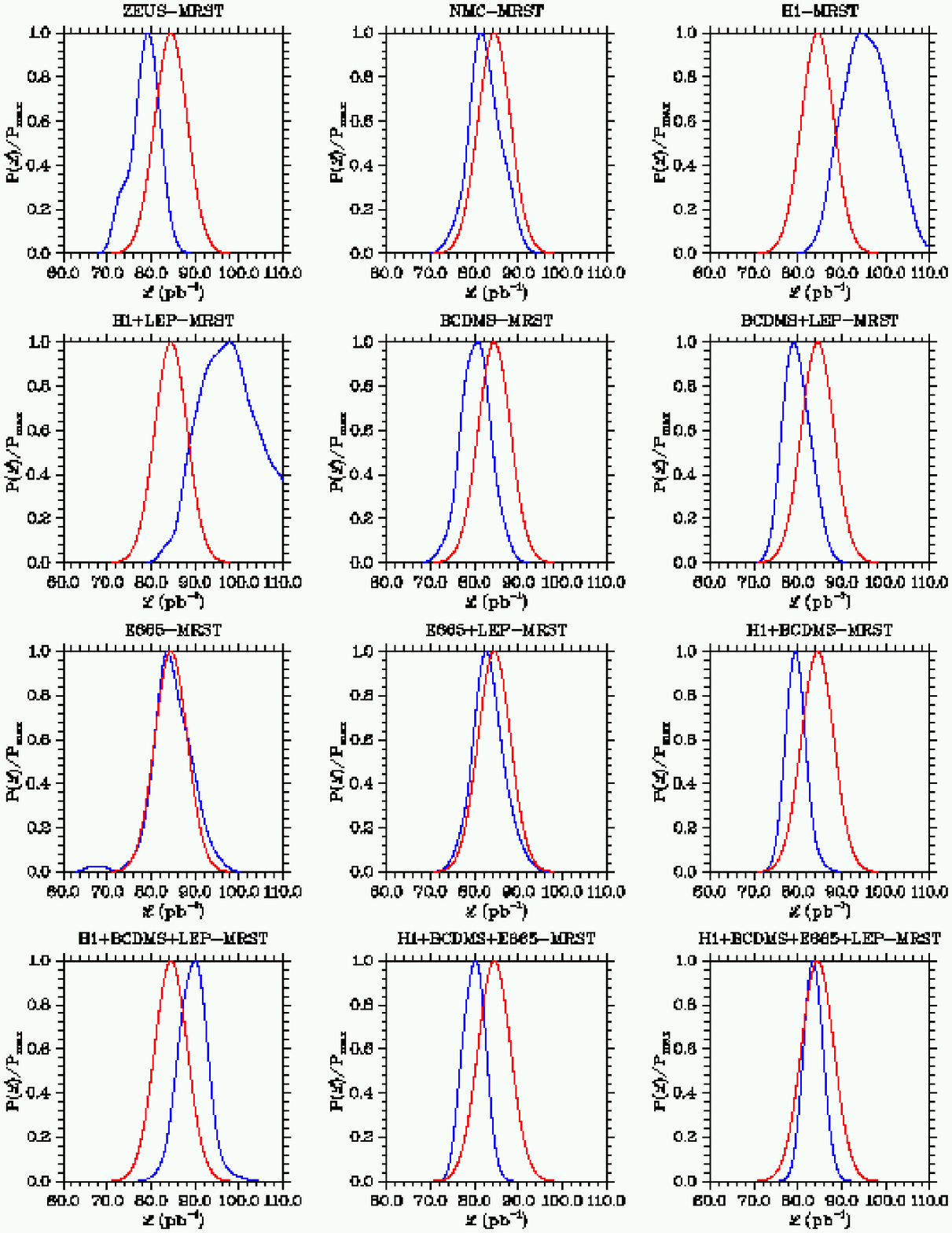}
\end{center}
\caption[]{The run 1b luminosity determination is shown.
In red is the D0 inelastic $p\bar p$ based determination,
while in blue is the luminosity probability density function
based on the \W and \Z event rates.}
\end{figure}

We could use the individual \W or \Z cross section to determine the luminosity. However,
such a method would not give us a luminosity independent measure of how well the
data describes the theoretical model. This is an important question as not all the
optimized sets might be correctly describing the hadron collider data.
The ratio $R$ of the \W and \Z cross sections gives us a luminosity independent
quantity and could function as a measure of the wellness of the particular \pdf
set to describe the data. This leads to the obvious method of deriving the 
luminosity from the correlated \W and \Z events. Using the D0 results of
table~1 we can determine experimental luminosity response function
\beq\label{lumiresponse}
P_{exp}^{luminosity}({\cal L}|\sigma_W,\sigma_Z,N_W,N_Z)=
\frac{1}{2\pi\sqrt{|C_{ij}|}}
\exp{\left(-\frac{1}{2}D_i\, C_{ij}^{-1}\, D_j\right)}
\eeq
where
\beq
D = \left(
            \begin{array}{c}
            {\cal L}\times\sigma_W-N_W \\
            {\cal L}\times\sigma_Z-N_Z 
            \end{array}
    \right)
\eeq
and $C_{ij}$ the error correlation matrix. 
The run 1a and run 1b one standard deviation 
ellipses together with the 100 prediction of 
each of the optimized \pdfs are shown in fig.~1
As is obvious from the figure the increased 
accuracy of run 1b has a dramatic impact
on the experimental result. The D0 one standard deviation
ellipses are show using the measured
luminosity and the luminosity uncertainty itself is incorporated in the contour.
Changing the value of the luminosity will move the ellipse over the green
line of constant ratio of the \W and \Z cross sections. 

Now we have to convert the results of fig.~1
into a luminosity measurement including the \pdf uncertainty.
For the remainder the experimental luminosity uncertainty is excluded from
the calculation of the correlation matrix in the experimental luminosity function
of eq.~\ref{lumiresponse}.
The \pdf probability function for the combined \W and \Z cross sections is in
the monte carlo approximation given by the scatter function
\beq\label{WZpdf}
P_{pdf}(\sigma_W,\sigma_Z)=\frac{1}{N}\sum_{i=1}^N 
\delta\left(\sigma_W-\sigma_W^{(i)}\right)\delta\left(\sigma_Z-\sigma_Z^{(i)}\right)
\eeq
where $\sigma_V^{(i)}=\sigma_V(\F_i)$. Note that this scatter function is shown in
fig.~1.
Using this \pdf probability function for the \W and \Z 
cross section together with the
experimental luminosity response function of eq.~\ref{lumiresponse}
we can construct the luminosity probability density function given the number of observed
\W, $N_W$, and \Z events, $N_Z$, 
\beqn\label{Bdef}
P_{pdf}({\cal L}|N_W,N_Z)\!\!\!\!&=&\!\!\!\!\int d\,\sigma_W\, d\,\sigma_Z\,
P_{exp}^{luminosity}({\cal L}|\sigma_W,\sigma_Z,N_W,N_Z)\times P_{pdf}(\sigma_W,\sigma_Z)\nonumber \\
&=&\!\!\!\!\frac{1}{N}\sum_i^N
P_{exp}^{luminosity}({\cal L}|\sigma_W^{(i)},\sigma_Z^{(i)},N_W,N_Z)\ ,
\eeqn
and the probability measure expressed in the confidence level
\beq
CL({\cal L}|N_W,N_Z)=\int d\,{\cal L}^\prime\ 
\Theta(P_{pdf}({\cal L}|N_W,N_Z)-P_{pdf}({\cal L}^\prime|N_W,N_Z))\times 
P_{pdf}({\cal L}^\prime|N_W,N_Z)\ .
\eeq
\begin{table}[t]\hspace{-1.5cm}\begin{center}
\begin{tabular}{|l||c|c||c|c|}\cline{2-5}
\multicolumn{1}{c|}{}&\multicolumn{2}{c||}{D0 1a} &\multicolumn{2}{c|}{D0 1b} \\ \cline{2-5}
\multicolumn{1}{c|}{}& ${\cal L}$ (pb$^{-1}$) & $CL^{exp}$ & 
                       ${\cal L}$ (pb$^{-1}$) & $CL^{exp}$ 
\\ \hline
$\sigma_{tot}(P\bar P)$ & 12.8+0.7-0.7 & - & 81.5+3.7-3.7 & - \\ \hline 
\scriptsize{ZEUS-MRST}              & 12.3+0.5-0.5 & 96 & 79.3+3.1-3.3 & {\red 4.6 (8.6)} \\
\scriptsize{NMC-MRST}               & 12.6+0.7-0.6 & 26 & 81.5+4.3-2.9 & 36 (29)\\
\scriptsize{H1-MRST}                & 14.2+1.0-0.9 & 75 & 94.4+6.6-4.5 & 41 (52)\\
\scriptsize{H1+LEP-MRST}            & 14.3+1.0-0.8 & 94 & 97.9+5.5-7.9 & 12 (26)\\
\scriptsize{BCDMS-MRST}             & 12.3+0.6-0.6 & 47 & 80.8+2.8-3.8 & 93 (93)\\
\scriptsize{BCDMS+LEP-MRST}         & 12.1+0.6-0.5 & 61 & 79.1+3.6-2.6 & 52 (66)\\
\scriptsize{E665-MRST}              & 12.8+0.9-0.7 & 51 & 83.7+5.3-3.1 & 80 (92)\\
\scriptsize{E665+LEP-MRST}          & 12.7+0.7-0.6 & 54 & 82.6+3.8-3.1 & 66 (72)\\
\scriptsize{H1+BCDMS-MRST}          & 12.0+0.5-0.5 & 94 & 79.5+2.3-2.3 & {\red 7.4} (12)\\
\scriptsize{H1+BCDMS+LEP-MRST}      & 13.6+0.7-0.7 & 56 & 90.0+2.8-3.7 & 71 (81)\\
\scriptsize{H1+BCDMS+E665-MRST}     & 12.3+0.5-0.5 & 54 & 80.3+2.3-2.8 & 67 (81)\\
\scriptsize{H1+BCDMS+E665+LEP-MRST} & 12.8+0.5-0.4 & 74 & 83.5+2.3-2.1 & 23 (34)\\
\hline\end{tabular}
\caption[]{The 31.73\%
confidence level intervals for the luminosity measurements based on the total $P\bar P$ cross sections
together with the determination based on the correlated \W and \Z measurement. Also shown is the
confidence level of the data describing the theory at the optimized luminosity,
$CL^{exp}\equiv CL(N_W^{exp},N_Z^{exp})$. Also indicated in brackets is the confidence level
using the next-to-next-to-leading order matrix elements.}
\end{center}\end{table}
Using $CL({\cal L}|N_W^{exp},N_Z^{exp})$ we calculate the 31.73\%
confidence level interval as an estimator of the luminosity
based on the experimental observations. The results are listed in table~6.
As can be seen the derived luminosities are very competitive with the traditional
determination used by D0.

However, the confidence level intervals give no indication how well the experiment is
described by the theory for the preferred luminosity. For this we have to calculate
the probability a repeat of the measurement gives a worse agreement with the theory at
the optimum luminosity. To do this we consider all possible outcomes of the experiment
and integrate the \pdf probability density function over the 
regions where the agreement with the theory is worse
\beqn
CL(N_W,N_Z)\!\!\!&=&\!\!\!\int d\,N_W^\prime\,d\,N_Z^\prime 
\Theta\left(\max_{\cal L}(P_{pdf}({\cal L}|N_W,N_Z))
           -\max_{\cal L}(P_{pdf}({\cal L}|N_W^\prime,N_Z^\prime))\right)\nonumber\\
&&\times \max_{\cal L}(P_{pdf}({\cal L}|N_W^\prime,N_Z^\prime))\ .
\eeqn
Note that the functional dependence is actually one dimensional as this confidence
level is scale invariant, i.e. $CL(N_W,N_Z)=CL(\kappa N_W,\kappa N_Z)$. In this sense
the luminosity independent confidence level defined here is equivalent to the more
traditional measure of agreement between experiment and theory of the ratio $R=N_W/N_Z$.
The results are listed in table~6. For run 1a all optimized \pdfs have a 
satisfactory agreement at the preferred luminosity. As the experimental uncertainties are
strongly reduced for the run 1b results the agreement between this experimental result
and the theory is more challenging. Even so most optimized \pdfs do very well. Also
indicated in the table for the run 1b results is the confidence level using the 
next-to-next-to-leading order matrix elements. It is useful to see the effect
of the truncation of the perturbative series for the hard matrix element. As is
clear the inclusion of higher order seems in general to increase the agreement between
experiment and theory. However, for a true estimate of the 
perturbative component of the uncertainty we need to
increase the order of the \pdfs evolution as well. 
Currently this is not possible and will have
to wait until all calculations have been completed~\cite{NeervenEvolve}.

\section{Using the measured luminosity.}

\begin{center}\begin{table}[t]\hspace{-1.0cm}
\begin{tabular}{|l||c|c||c|c|}\cline{2-5}
\multicolumn{1}{c|}{}& $\sigma_{t\bar t}^{pdf}$ (pb) & $\sigma_{t\bar t}^{obs}$ (pb)   
& $N_{t\bar t}^{pdf}$ & $CL(N_{t\bar t}^{obs})$\\ \hline
MRS99 & 5.17 & - & - & - \\
CTEQ5M & 5.39 & - & - & - \\
\hline\hline
\scriptsize{ZEUS-MRST}              & 3.15+0.71-0.26 & 5.87+0.26-0.22 & 242+40-10 & {\red 3.8} \\
\scriptsize{NMC-MRST}               & 5.66+1.32-1.20 & 5.71+0.20-0.29 & 415+41-20 & 72.6 \\
\scriptsize{H1-MRST}                & 4.09+1.05-0.52 & 4.93+0.25-0.32 & 410+20-66 & 51.2 \\
\scriptsize{H1+LEP-MRST}            & 5.33+0.19-1.53 & 4.75+0.43-0.27 & 452+28-117 & 65.1 \\
\scriptsize{BCDMS-MRST}             & 4.21+0.27-0.27 & 5.76+0.28-0.20 & 338+22-18 & 27.1 \\
\scriptsize{BCDMS+LEP-MRST}         & 5.29+0.29-0.60 & 5.88+0.20-0.25 & 420+26-26 & 64.6 \\
\scriptsize{E665-MRST}              & 4.05+2.45-0.36 & 5.56+0.22-0.32 & 345+134-16 & 54.0 \\
\scriptsize{E665+LEP-MRST}          & 4.89+2.74-2.44 & 5.63+0.22-0.24 & 400+96-145 & 46.6 \\
\scriptsize{H1+BCDMS-MRST}          & 4.48+0.13-0.16 & 5.85+0.17-0.16 & 356+12-12 & 32.8 \\
\scriptsize{H1+BCDMS+LEP-MRST}      & 4.25+0.13-0.22 & 5.17+0.22-0.15 & 379+14-13 & 43.1 \\
\scriptsize{H1+BCDMS+E665-MRST}     & 4.95+0.42-0.29 & 5.80+0.21-0.15 & 401+17-16 & 53.0 \\
\scriptsize{H1+BCDMS+E665+LEP-MRST} & 4.35+0.21-0.15 & 5.57+0.15-0.14 & 364+14-11 & 37.5 \\
\hline
\end{tabular}
\caption[]{The collection of top quark pair predictions using the optimized \pdf sets.
The first column is the 31.73\%
confidence level interval using eqs.~\ref{theoryH} and~\ref{CLdef}.
The second column is the measured D0 cross section using eq.~\ref{ttcross}.
The third column is the expected number of top quark pair events using eq.~\ref{ttN}.
The last column is the confidence level of the observed number of top quark pair events
using the probability density of eq.~\ref{ttNexp}.}
\end{table}\end{center}
The luminosity determined in the previous section using next-to-leading order
perturbative QCD can be used for the other data in the experiment. 
We will take as an example the topquark pair production. In the first
column of table~7 we show the 31.73\%
confidence level theory predictions including the \pdf
uncertainties using eqs.~\ref{theoryH} and~\ref{CLdef}
with $\Delta=0.1$ pb. As can be seen the \pdf
uncertainties on the topquark pair cross section are substantial.
The published D0 run 1b topquark pair event rate is based on an enlarged data set
with an integrated luminosity of ${\cal L} = 125$ pb$^{-1}$. To estimate the efficiency
corrected number of topquark pair events 
for the \W and \Z sample 
we simply scale the luminosity down to the
quoted D0 run 1b integrated luminosity of 84.5 pb$^{-1}$.
This gives a number of topquark pair events of $N_{t\bar t}^{obs} = 465\pm 150$. Using this
number of observed events
we can derive the measured cross sections by taking the ratio of the number
of observed events over the luminosity measurements. The probability density function
is given by eq.~\ref{Bdef} with the substitution ${\cal L}=N_{t\bar t}/\sigma_{t\bar t}$
\beqn\label{ttcross}
P_{pdf}(\sigma_{t\bar t}|N_{t\bar t},N_W,N_Z)\!\!\!&=&\!\!\!\frac{1}{N}\sum_{i=1}^N
P_{exp}^{luminosity}(N_{t\bar t}/\sigma_{t\bar t}|\sigma_W^{(i)},\sigma_Z^{(i)},N_W,N_Z)\nonumber\\
&=&\!\!\!P_{pdf}(N_{t\bar t}/\sigma_{t\bar t}|N_W,N_Z)\ .
\eeqn
The 31.73\%
confidence level interval is given in the second column of table~7. Note that 
only the uncertainty induced through the luminosity uncertainty is included.
The actual uncertainty on the number of observed topquark pair events is not as it
would overwhelm the luminosity uncertainty for the current results. To include the
experimental uncertainty in the topquark pair cross section we have to use
the experimental response function density $P_{exp}(N_{t\bar t}^{observed}|N_{t\bar t}^{nature})$
(i.e. the detector uncertainty) to get the topquark pair cross section probability function
\beq
P_{obs}(\sigma_{t\bar t}|N_{t\bar t},N_W,N_Z)=\int d\,N_{t\bar t}^{nat}
P_{exp}(N_{t\bar t}|N_{t\bar t}^{nat})\times P_{pdf}(N_{t\bar t}^{nat}/\sigma_{t\bar t}|N_W,N_Z)\ .
\eeq
The advantage of this way of using the luminosity is that one can compare
the derived topquark pair cross section with other experiments.

When comparing directly with the theory one can use the luminosity correlation between
the vectorboson production and the topquark pair production to further reduce the 
luminosity uncertainty.
This we do by prediction the expected number of topquark pairs given the number of \W and \Z
events. The resulting formula is closely related to eq.~\ref{ttcross}, however now the luminosity
substitution is inside the monte carlo summation over \pdfs as $\sigma_{t\bar t}$ now is dependent
on the \pdf
\beq\label{ttN}
P_{pdf}(N_{t\bar t}|N_W,N_Z)=\frac{1}{N}\sum_{i=1}^N
P_{exp}^{luminosity}(N_{t\bar t}/\sigma_{t\bar t}^{(i)}|\sigma_W^{(i)},\sigma_Z^{(i)},N_W,N_Z)\ ,
\eeq
where the triplet $(\sigma_W^{(i)},\sigma_Z^{(i)},\sigma_{t\bar t}^{(i)})$ are the next-to-leading
order predictions using \pdf \F$^{(i)}$ out of the optimized set. 
The results are shown in the third column of table~7.

Using the experimental response
function we get the smooth prediction for the probability density function of observing $N_{t\bar t}$
topquark pair events given $N_W$ and $N_Z$ vectorboson events, now including the experimental
detector uncertainties
\beq\label{ttNexp}
P_{obs}(N_{t\bar t}|N_W,N_Z)=\int d\,N_{t\bar t}^{nat}
P_{pdf}(N^{nat}_{t\bar t}|N_W,N_Z)\times P_{exp}(N_{t\bar t}|N_{t\bar t}^{nat})\ .
\eeq
By converting this probability density to a confidence level probability one can
calculate the likelyhood the observed number of topquark pair events agrees with
the theory predictions. The confidence level
for the ``observed'' number of topquark pair events is shown in column 4
of table~7. Note the 32\%
exerimental uncertainty is now included in the estimate.
 
\section{Conclusions}

The luminosity determination using the \W and \Z event rates can easily compete
with the traditional methods with respect to accuracy. An added feature is that 
when comparing observables to the theory the luminosity uncertainty partly cancels 
because the observable dependence on the \pdfs is correlated to the \W and \Z
dependence. This leads to more accurate comparisons between theory and experimental
result. The traditional luminosity determination
offers no such correlations as it is not based on perturbatively calculable
processes. Also, by including additional measurements in 
the \pdf optimalization we can systematically
improve the luminosity uncertainty to a level required by the physics of the
TEVATRON run 2 or the LHC. Furthermore, the method can be extended to 
next-to-next-to-leading order once the required calculations are completed
resulting in an excellent control of theoretical uncertainties.

Using the preferred {\tt H1+BCDMS+E665-MRST} \pdf set,
which includes data from three mutually consistant experiments,
we find a predicted \W cross
section of $2.44\pm 0.07$ nb where the uncertainty is due to the \pdfsb. Comparing
with the D0 measured cross section in run 1b this leads to a confidence level of 26\%.
Similar, the predicted \Z cross section of $0.232+0.007-0.006$ gives a run 1b 
confidence level of 34\%.
From this we can conclude the optimized \pdf set describes the collider physics well
in parton fraction range relevant for the vectorboson physics. This means we can 
confidently continue to determine the run 1b D0 integrated luminosity for this sample.
We find an integrated luminosity of $80.3+2.3-2.8$ pb$^{-1}$ with a maximized confidence
level of 67\%,
reflecting the fact that the correlated \W and \Z data is well described by the \pdf set.
Using the measured luminosity we can continue to predict the number of observed  
topquark pair events. Including the \pdf and luminosity uncertainty we expect 
$401+17-16$. Comparing to the measured (but scaled) D0 run 1b measurement one 
finds a confidence level of 53\%.

The method described in this paper can at minimum be used as an check on the traditional
luminosity measurement. However, given its potential better accuracy and partial cancellation
of the correlated luminosity uncertainties one can contemplate replacing the traditional 
method in future experiments. The TEVATRON run II results will 
offer an excellent testing ground for these ideas.

\end{document}